\documentclass{article}
\usepackage{spconf,amsmath,graphicx,hyperref}
\usepackage{float}
\usepackage{caption}
\usepackage{multirow,multicol,makecell, ragged2e}

\usepackage{enumitem}
\setlist{nosep, leftmargin=14pt}

\usepackage{mwe} 
\usepackage[ruled, linesnumbered]{algorithm2e} 
\DontPrintSemicolon
\usepackage{amsfonts} 
\usepackage{caption}
\usepackage{lipsum}
\usepackage{xcolor}
\usepackage{todonotes}
\usepackage{booktabs}
\usepackage{amssymb}

\title{Sliding Window FastEdit: A Framework for Lesion Annotation in Whole-body PET Images}

\name{
    \begin{tabular}{@{}c@{}}
$^*$Matthias Hadlich$^{1}$ \quad
$^*$Zdravko Marinov$^{1,2}$  \quad 
Moon Kim$^{3}$ \\
Enrico Nasca$^{3}$ \quad
Jens Kleesiek$^3$ \quad 
Rainer Stiefelhagen$^1$
\end{tabular}
}
\address{
$^1$ Institute for Anthropomatics \& Robotics (IAR), Karlsruhe Institute of Technology, Germany 
\\
$^2$ HIDSS4Health - Helmholtz Information and Data Science School for Health, Karlsruhe, Germany
\\
$^3$ Institute for AI in Medicine (IKIM), University Hospital Essen, Germany
}

\setlipsum{%
  par-before = \begingroup\color{lightgray},
  par-after = \endgroup
}

\begin{document}
%
\maketitle
\begin{abstract}
Deep learning has revolutionized the accurate segmentation of diseases in medical imaging. 
However, achieving such results requires training with numerous manual voxel annotations. 
This requirement presents a challenge for whole-body Positron Emission Tomography (PET) imaging, where lesions are scattered throughout the body.
To tackle this problem, we introduce \textsc{SW-FastEdit} -- an interactive segmentation framework that accelerates the labeling by utilizing only a few user clicks instead of voxelwise annotations. While prior interactive models crop or resize PET volumes due to memory constraints, we use the complete volume with our sliding window-based interactive scheme. 
Our model outperforms existing non-sliding window interactive models on the AutoPET dataset and generalizes to the previously unseen HECKTOR dataset. A user study revealed that annotators achieve high-quality predictions with only 10 click iterations and a low perceived NASA-TLX workload. Our framework is implemented using MONAI Label and is available  \href{https://github.com/matt3o/AutoPET2-Submission/}{here}.


\end{abstract}
\makeatletter{\renewcommand*{\@makefnmark}{}
\footnotetext{* shared first author}\makeatother}
\section{Introduction}

\begin{figure}[t!]
    \centering
    \includegraphics[width=\linewidth]{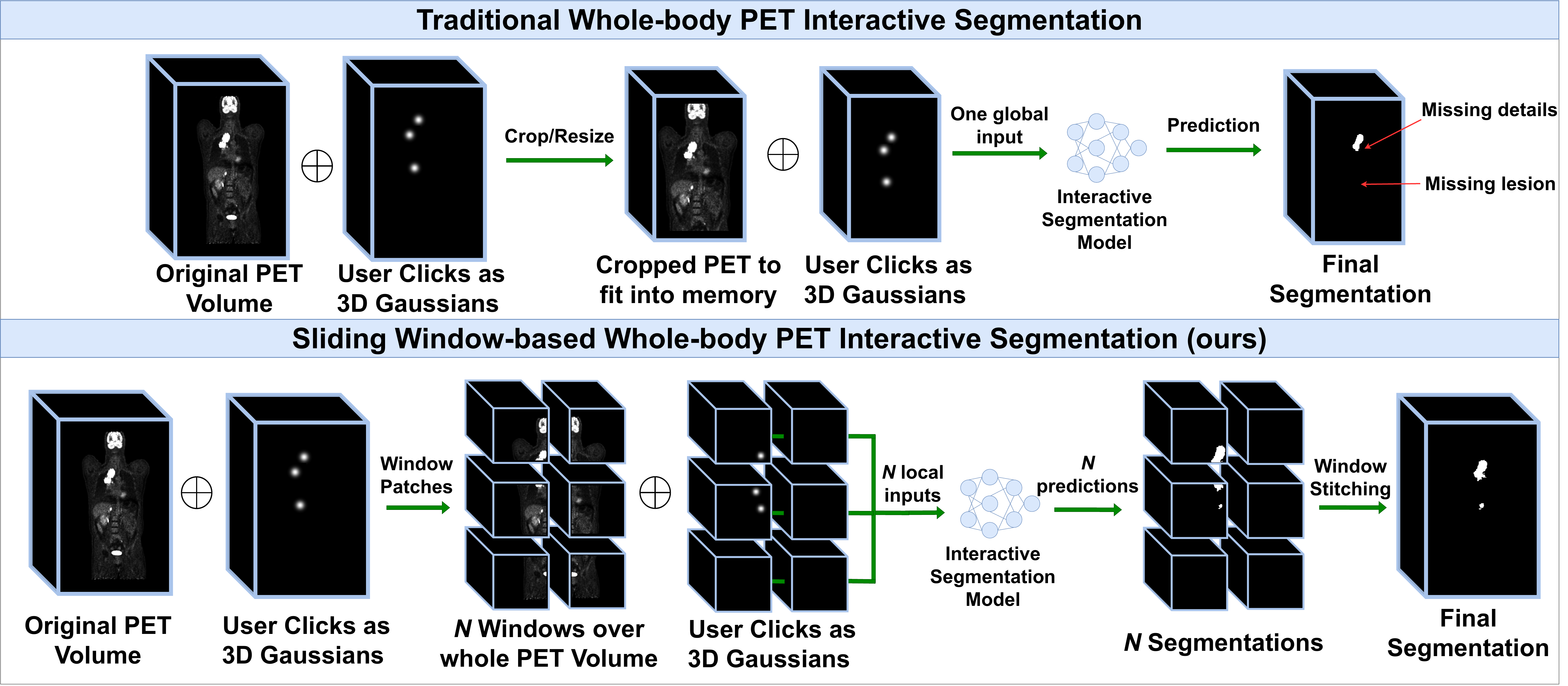}
    \caption{Traditional interactive models need to crop or resize the whole-body PET due to memory constraints, potentially missing small lesions, and detailed boundaries (top row).  
    Our sliding window approach partitions the volume into patches and simulates corrective clicks in poorly segmented patches. This allows our model to utilize the full-resolution PET volume and improve upon traditional methods (bottom row).}
    \label{fig:teaser}
\end{figure}

Supervised Deep Learning (DL) models have achieved remarkable performance in visual tasks such as classification, object recognition, and semantic segmentation, owing to the widespread availability of extensive manually labeled datasets~\cite{russakovsky2015imagenet, kirillov2023segment}. However, annotating lesions in whole-body volumetric PET data presents unique challenges. This is primarily due to the diverse range of lesion shapes and locations as well as the complex three-dimensional structure and size of the whole-body PET images \cite{heiliger2022autopet, bendazzoli2022priornet, Marinov_2023_ICCV} leading to tedious annotation times of up to 60 minutes per volume ~\cite{gatidis2023autopet}.

Interactive segmentation models address these concerns by expediting annotation while enhancing label accuracy. 
Interactive models employ user interactions, such as clicks or scribbles, to guide the model toward the region of interest, and iteratively refine its prediction with new interactions until the annotator is satisfied with their quality \cite{marinov2023guiding, diaz2022deepedit, 10230334}. This efficient approach significantly reduces voxel-wise annotations to a few user interactions, decreasing the annotation time for a whole-body PET image to a few minutes \cite{10230334}. This efficiency has led to the growing popularity of interactive models in the realm of 3D segmentation for medical image analysis.

\begin{figure*}[!t]
    \centering
    \includegraphics[width=0.7\textwidth]{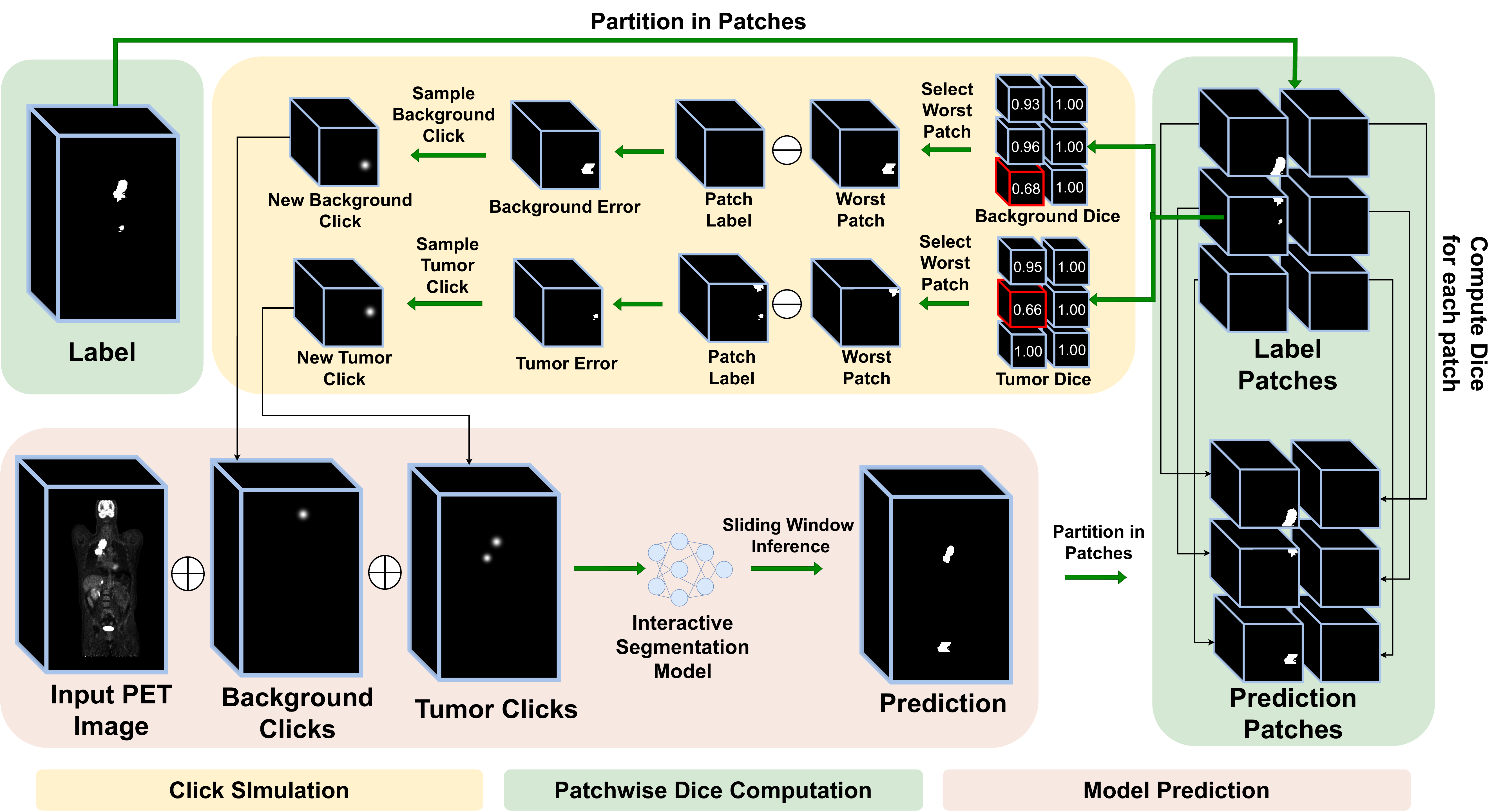}
    \caption{Local patch-wise corrections. After the model predicts a tumor mask (red box), the prediction and tumor label are partitioned into patches, and the Dice score for each patch is computed (green box). Then, the patch with the worst Dice score is selected for both the "tumor" and "background" classes, and a click is sampled from the error within each patch (yellow box).}
    \label{fig:patch_based_guidance}
\end{figure*}

Prior interactive models achieve impressive results on Magnetic Resonance Imaging (MRI) \cite{wang2018deepigeos}, Computer Tomography (CT) \cite{diaz2022deepedit}, and PET/CT \cite{10230334} images. In cases with large volumes, such as a whole-body PET/CT, interactive methods opt to resize \cite{10230334} or crop the volume \cite{diaz2022deepedit, marinov2023guiding} due to memory constraints. However, resizing results in resolution loss, and cropping risks discarding valid labels beyond the crop.

To tackle these challenges, we introduce \textsc{SW-FastEdit} -- a method that smoothly blends sliding window inference into the training and evaluation of interactive models. This process entails dividing volumes into smaller patches, allowing the model to segment them individually, and subsequently assembling them into the final segmentation, as seen in Figure \ref{fig:teaser}, and eliminates the need for resizing or cropping. \textsc{SW-FastEdit} builds on top of the GtG model \cite{marinov2023guiding}, originally based on the DeepEdit MONAI Label codebase \cite{diaz2022deepedit}, by applying sliding window strategies during both training and evaluation. \textsc{SW-FastEdit} is model-agnostic and can be used with any deep learning architecture. This marks the first interactive segmentation method based on sliding window inference, possibly due to the non-differentiable nature of the sliding window. Our contributions are summarized as follows:
\begin{itemize}
\item We integrate the sliding window inference into the interactive segmentation paradigm.
\item We extensively evaluate our method, covering: 1) diverse click simulation strategies; 2) non-interactive pretraining; 3) generalization to another PET/CT dataset; 4) comparison to prior work; 5) and a user study.
\item We provide publicly available code and trained weights for PET/CT lesion segmentation via the MONAI Label framework \cite{MONAILabel}.
\end{itemize}



\section{Method}
Our proposed \textsc{SW-FastEdit} utilizes sliding window inference to partition whole-body PET images into patches and infers predictions for each patch. As sliding window-based interactive models remain underexplored, we investigate various training and evaluation strategies, along with essential sliding window hyperparameters to enhance performance.

\subsection{Click Simulation}
Our interactive model training comprises two key steps: 1) a click generation strategy; 2) and a stopping criterion. We explore multiple variants for these steps to optimize the training of our sliding window-based approach.

\textbf{Click Generation Strategies.} Click generation involves how clicks are created and integrated into the model's input. Previous work has focused on two simulation paradigms: 1) non-corrective \cite{wang2018deepigeos}; 2) and corrective click simulation \cite{sakinis2019interactive}. Non-corrective clicks are generated either via fixed rules, such as placing a click at the center of the largest connected label component \cite{zhang2021interactive}, or via random sampling of clicks from the ground truth mask \cite{wang2018deepigeos}. In both cases, all clicks are generated in one step, combined with the image, and processed in a single model prediction. In contrast, corrective clicks are introduced over multiple prediction steps, with new clicks generated in regions where the model has made errors. In this paradigm, new clicks can also be placed using fixed rules, such as in the center of the largest error region \cite{liu2022isegformer}, or through sampling from the error area \cite{sakinis2019interactive}. This cycle of prediction steps concludes upon meeting a stopping criterion.

\textbf{Stopping Criterion.} To prevent infinite loops, corrective click strategies utilize a stopping criterion. Many previous methods employ a fixed number of prediction steps, denoted as $N_\text{max}$, and stop after reaching this limit \cite{diaz2022deepedit, liu2022isegformer, sakinis2019interactive}. We additionally experiment with stopping at each prediction step with a probability $p \in [0,1]$. As a result, the number of steps varies for different training samples, reducing the model's reliance on clicks to produce high-quality segmentations \cite{marinov2023guiding}. We introduce an additional criterion for stopping, which triggers when the model achieves a satisfactory Dice score of $\text{Dice}_\text{max}$ on the sample. This mirrors the way radiologists annotate samples, halting the process when they are content with the label quality. We also investigate the potential benefits of combining multiple stopping criteria during training.

\subsection{Inference Strategies}
We investigate two methods for inference with our model: 1) global corrections; 2) and local patch-wise corrections. Global corrections involve sampling new clicks based on the errors across the entire volume. On the other hand, local patch-wise corrections enable us to sample clicks from the patch window with the lowest Dice score, simulating an annotator correcting the poorly segmented local regions. This process is illustrated in Fig. \ref{fig:patch_based_guidance}. It is important to note that these two inference strategies and the stopping criteria are only applicable when simulating corrective clicks.

\subsection{Training Details}
\textbf{Data Pre-processing.} We use the AutoPET dataset \cite{gatidis2023autopet} for training and validation containing whole-body PET/CT scans of patients with lung cancer, melanoma, or lymphoma. We only use PET volumes of unhealthy patients and we split them into 396 training and 105 test samples without containing the same patient in both splits. We normalize the PET volumes and keep values between the 0.05th and 99.95th percentiles of the batch. We randomly crop a subvolume with a size of $224 \times 224 \times 224$ voxels and a probability $p_{\text{tumor}}=0.6$ to be centered around a tumor and $p_{\text{bg}}=0.4$ around a non-tumor voxel. We apply random flipping with a probability of $p_\text{flip}=0.1$ for each 3D axis and random rotation with $p_{\text{rot}}=0.1$ for each 3D axis. We then feed the transformed input to our sliding window-based model with a window size of $128 \times 128 \times 128$ and a window overlap of $25\%$ with Gaussian weighting.

\textbf{Click Simulation and Inference.} We train all our models using global corrections and investigate two click generation strategies along with five stopping criteria for simulating clicks. These are detailed in Table \ref{tab:click_inference_strategies}. We encode clicks as 3D spheres with a radius of one voxel ($\sigma=1$), and concatenate them to the PET images as an additional channel, as depicted in Fig. \ref{fig:teaser}. During the first prediction step, this channel is empty, and with each simulated corrective click, a new sphere is generated within this input channel. Note that we simulate a "tumor" and "background" click in each prediction step to correct under- and oversegmentation respectively by sampling from the distance map of the error regions as in \cite{diaz2022deepedit}. This leads to two additional input channels. We set $p=0.5$ and $N_\text{max}=10$ as the default values from \cite{marinov2023guiding}.

\textbf{Model Optimization.} As sliding window inference is not differentiable by default we utilize MONAI's \cite{cardoso2022monai} differentiable implementation of the \textsc{SlidingWindowInferer} for the first time for interactive segmentation, where predictions on overlapping windows are weighted based on the error from each prediction and then averaged so that gradients are linearly combined. We further utilize a cosine annealing learning rate scheduler with an initial learning rate of 1e-4 and train our models for 200 epochs. We utilize the MONAI DynUNet backbone with six encoder-decoder levels.

\section{Experiments and Results}
We conduct quantitative experiments on the AutoPET \cite{gatidis2023autopet} and HECKTOR \cite{oreiller2022head} PET/CT datasets and a user study with one radiologist, one medical doctor, and two medical students to showcase the usability of our model.

\subsection{Quantitative Experiments}

\textbf{Click Simulation and Inference Strategies.} Table \ref{tab:click_inference_strategies} displays the results when using various click generation strategies. The corrective paradigm consistently outperforms the non-corrective, demonstrating a substantial improvement of over $34\%$ across all cases. Notably, incorporating $p=50\%$ as a stopping criterion leads to a decline in performance while leveraging $\text{Dice}_\text{max}=90\%$ enhances it. We validate models with both inference strategies to assess their robustness to various annotation styles, e.g., an annotator focusing on global or local errors. The best results, indicated in \textbf{bold}, are achieved with local patch-wise inference. Note that the second row corresponds to a sliding window version of DeepEdit \cite{diaz2022deepedit}.

\begin{table}[!h]
\scalebox{0.7}{
    \begin{tabular}{l|c|c}
        \toprule
        \multirow{2}{*}{Click Generation Strategy (Training)} & \multicolumn{2}{c}{Inference Strategy (Validation)} \\ \cline{2-3}
        {} & Global & Local Patch-wise \\ \hline
        Non-corrective (10 clicks) & $47.00\%$ & $47.00\%$ \\
        Corrective: $N_\text{max}=10$ (DeepEdit \cite{diaz2022deepedit}) & $84.90\%$ & $82.02\%$ \\
        Corrective: $N_\text{max}=10$, $p=50\%$ & $81.07\%$ & $81.23\%$ \\
        Corrective: $N_\text{max}=10$, $\text{Dice}_\text{max}=90\%$  & \textbf{85.34}$\%$  & \textbf{85.50$\%$}  \\
        Corrective: $N_\text{max}=10$, $p=50\%$, $\text{Dice}_\text{max}=90\%$ & $82.54\%$  & $82.93\%$ \\

        \bottomrule
    \end{tabular}}
    \caption{Results (Dice score) for \textsc{SW-FastEdit} when using different click generation strategies during training and various inference strategies during validation. During validation, we always simulate exactly 10 clicks for each sample.}
    \label{tab:click_inference_strategies}
\end{table}

\textbf{Non-interactive Pre-training.} We explore the impact of initial pre-training without interactions, i.e., with empty click channels. Table \ref{tab:retraining} reveals that our top-performing model from Table \ref{tab:click_inference_strategies} exhibits poor performance in the absence of clicks (Dice@0$=24.47\%$), particularly when compared to a model trained without any clicks (Dice@$0=73.04\%$). However, pre-training a model without clicks for 400 epochs, followed by 200 epochs with clicks, significantly improves the initial prediction (Dice@$0=68.03\%$) while competing with the performance of our best model (Dice@$10=84.79\%$). 

\begin{table}[h!]
    \centering
        \begin{tabular}{l|c|c|c}
            \toprule
            400 epochs w/o clicks & {} & \checkmark & \checkmark \\
            200 epochs with clicks & \checkmark & {} & \checkmark \\
            \hline
            Dice@0 & $24.47\%$ & $73.04\%$ & $68.03\%$\\
            Dice@10 & $85.50\%$ & - & $84.79\%$ \\

            \bottomrule    
        \end{tabular}
        \caption{Results from non-interactive pre-training with local inference. Dice@X denotes the Dice score after X clicks.}
            \label{tab:retraining}
\end{table}

\begin{table}[h!]
    \centering
    \scalebox{0.8}{
    \begin{tabular}{l|cc}
    \toprule
    {} & \textsc{SW-FastEdit} & DeepEdit \cite{diaz2022deepedit}  \\ \cline{2-3}
    Inference (global) & sliding window & standard  \\
    Dice@10 ($224^3$ crop) & \textbf{85.55}$\%$ & $84.34\%$ \\
    Dice@10 (full volumes) & $84.90\%$ & does not fit on 48 GB GPU \\
    \bottomrule
    \end{tabular}}
    \caption{Comparison between the sliding window and standard inference. Dice@X is the Dice score after X clicks.}
    \label{tab:dice_score}
\end{table}

\textbf{Comparison with Non-Sliding Window Inference:} We compare \textsc{SW-FastEdit} trained with $N_\text{max}=10$ to the non-sliding window-based DeepEdit \cite{diaz2022deepedit}, also originally trained with $N_\text{max}=10$. We evaluate both models on $224 \times 224 \times 224$ center crops of the validation set to accommodate DeepEdit's challenge in handling large whole-body PET/CT volumes without resizing. This adjustment results in a shift from our initial $84.90\%$ Dice score to $85.55\%$ due to changes in the tumor/background ratio induced by cropping. Table \ref{tab:dice_score} shows that \textsc{SW-FastEdit} outperforms DeepEdit \cite{diaz2022deepedit} on the center crop and is also able to process the full volumes.

\textbf{Generalization to Unseen Data.} We tested our top model from Table \ref{tab:click_inference_strategies} on the HEad and NeCK TumOR (HECKTOR) \cite{oreiller2022head} PET/CT dataset without any fine-tuning. For the evaluation, we combine primary tumors and nodal tumors in a single "tumor" category. Using 0 clicks leads to Dice@$0 =3.59\%$, likely due to the domain shift. However, utilizing corrective clicks for the tumor and background significantly boosts the performance to Dice@$10=40.77\%$. This highlights the potential generalization of our interactive model, even when faced with previously unseen data.

\subsection{User Study}

We conducted a user annotation study with one radiologist, one medical doctor, and two medical students. The task was to annotate as many validation volumes as possible within a time limit of 80 minutes. The annotators were instructed to perform the following loop (1)-(3) \textbf{exactly} 10 times: (1) predict with our model; (2) add one tumor click; (3) add one background click. During steps (2) and (3), annotators corrected areas where our model had missegmented in step (1). Our user study employed the non-interactively pretrained model detailed in Table \ref{tab:retraining}. We assessed the user study using the Dice and Normalized Surface Dice (NSD) metrics, as well as the perceived NASA-TLX workload.

Table \ref{tab:annotator_statistics} presents the user study results. The simulated user achieves a Dice score of 78.50\% and NSD of 39.17\% with 10 clicks. Note that the simulated user always simulates valid clicks since he has access to the ground truth labels. The best annotator (A1) achieves a slightly lower Dice and NSD than the simulated user, however, his results are significantly better than the non-interactive model. This shows that an interactive model can deliver much higher quality results than non-interactive models alone. The time per annotation varies between 6-8 minutes per annotator and volume, compared to up to 60 minutes when manually annotating them \cite{gatidis2023autopet}.



\begin{table}[t]
   
    \resizebox{\columnwidth}{!}{%
    \begin{tabular}{l|c|c|c}
    \toprule
    Annotator & \#Volumes & Dice@10 & NSD@10\\ \hline
    A1 (medical doctor) & $12$ & \textbf{72.49}$\%$ $\pm$ $18.66\%$ & \textbf{39.17}$\%$ $\pm$ $24.63\%$ \\ 
    A2 (medical student) & $11$ & $64.66\%$ $\pm$ $23.13\%$ & $34.50\%$ $\pm$ $26.38\%$ \\ 
    A3 (radiologist) & $13$ & $67.72\%$ $\pm$ $21.00\%$ & $36.19\%$ $\pm$ $24.01\%$ \\ 
    A4 (medical student) & $10$ & $65.65\%$ $\pm$ $26.24\%$ & $34.33\%$ $\pm$ $27.14\%$   \\ \hline
    Simulated user@0  & $10$ & $51.43\%$ $\pm$ $25.21\%$ & $25.35\%$ $\pm$ $13.31\%$ \\ 
    Simulated user@10  & $10$ & \textbf{78.50}$\%$ $\pm$ $14.96\%$ & \textbf{45.14}$\%$ $\pm$ $22.99\%$ \\ \hline
    Non-interactive model & $10$ & $61.69\%$ $\pm$ $20.53\%$ & $32.85\%$ $\pm$ $17.69\%$ \\ \bottomrule
    \end{tabular}
    }
     \caption{Results from the user study. The metrics presented in the table correspond exclusively to the 10 volumes that all annotators reviewed in common, i.e. the 10 volumes of A4.}
    \label{tab:annotator_statistics}
\end{table}

\textbf{NASA-TLX and Questionnaire.} Following the labeling, we asked the annotators to fill out a NASA-TLX form, three Likert-scale questions, and one open question for feedback. The annotators ranked the mental (3.5/10), physical (2/10), and temporal demand (3.3/10), as well as the effort (4.3/10) and frustration (2.3/10) on average as low. They also rated their performance on the task relatively high (6.5/10). Additionally, the annotators rated 10 click iterations as sufficient (6.6/10), background clicks as necessary (6.5/10), and that \textsc{SW-FastEdit} speeds up the annotation (7.5/10), with A3 commenting: "majority of the cases can be annotated with only 3-4 updates, which is really great. The process was much faster than annotating PET images from scratch". Overall the feedback is positive and the annotators saw potential in applying \textsc{SW-FastEdit} in their annotation workflow.

\section{Discussion and Conclusion}
\textsc{SW-FastEdit} shows that incorporating sliding window inference enhances the performance of previous non-sliding window models, efficiently handling large PET/CT volumes without requiring resizing or cropping. Our approach demonstrates promising generalization to unseen data, with a user study indicating a low perceived NASA-TLX workload with medical experts expressing favorable opinions and indicating a willingness to use it. In our future work, we plan to assess the applicability of \textsc{SW-FastEdit} across diverse imaging modalities and segmentation tasks, examining the generalizability of the sliding window approach. Additionally, we aim to explore its utility in a multimodal setting, with a specific focus on tasks like PET/CT segmentation.



\section{Compliance with Ethical Standards}

This research study was conducted retrospectively using human subject data made available in open access. Ethical approval was *not* required as confirmed by the license attached with the open access data.
\section{Acknowledgments}
\label{sec:acknowledgments}

The user study was done in collaboration with the Annotation Lab Essen (\url{https://annotationlab.ikim.nrw/}). The present contribution is supported by the Helmholtz Association under the joint research school “HIDSS4Health – Helmholtz Information and Data Science School for Health“.

\bibliographystyle{IEEEbib}
\bibliography{strings,refs}

\end{document}